\documentstyle[aps,multicol,epsf]{revtex}

\begin{document}
\draft
\title{modeling study on the validity of a possibly 
simplified representation of proteins}
\author{Jun Wang and Wei Wang}
\address{National Laboratory of Solid State Microstructure
and Physics Department, Nanjing University, Nanjing 210093, China}
\date{\today}
\maketitle

\begin{abstract}
The folding characteristics of sequences reduced with a possibly 
simplified representation of five types of residues are shown to 
be similar to their original ones with the natural set of residues 
(20 types or 20 letters). The reduced sequences have a good 
foldability and fold to the same native structure of their optimized 
original ones. A large ground state gap for the native structure 
shows the thermodynamic stability of the reduced sequences. 
The general validity of such a five-letter reduction is further 
studied via the correlation between the reduced sequences and the 
original ones. As a comparison, a reduction with two letters is found 
not to reproduce the native structure of the original sequences due 
to its homopolymeric features. 
\end{abstract}

\widetext

\begin{multicols}{2}
\narrowtext

\section{INTRODUCTION}
Protein folding is a well known complicated and highly cooperative 
dynamic process due to the heterogeneity in proteins (e.g., see 
Refs.\cite{Books,Dillchan,Wolynes0} and references therein). Much effort 
has been made by considering minimalist models with a few types of 
amino acid residues to simplify the natural set of residues (of 20 
types) for better physical understanding 
\cite{HP,Wolynes1,Lihao1,Lihao2,enum1,TK1} and practical 
purposes \cite{exper1,exper2,Baker}. In these models the compositions 
are much simpler than the real ones. The simplest reduction is the 
so-called hydrophobic and polar ones with only two groups of 
residues (each group including some types of residues) by considering 
the main driving force, the hydrophobicity 
\cite{HP,Wolynes1,Lihao1,Lihao2,enum1,TK1}. Furthermore, 
these two groups actually are simplified as two effective monomers 
or letters, namely H and P, which is known as the HP model \cite{HP}. 
The studies of such a model enable people to understand some fundamental 
physics and mechanism of protein folding. However, as argued in a number 
of studies (see \cite{sh1} and reference therein), the HP model may 
be too simple and lacks enough consideration on the heterogeneity and the 
complexity of the natural set of residues, such as the interactions 
between the residues \cite{Wolynes1,Lihao1,WW}. Moreover, the minimal 
sets of residues for protein design suggested by biochemical experiments 
\cite{exper1,exper2,Baker} seem unfavorable to those with only two types 
of residues since a small number of types obviously introduces the 
homopolymeric degeneracy. What is the suitable simplification for natural 
proteins, or how many types of residues are necessary for reproducing 
some useful structures and for a simplified representation of protein 
sequence characteristics? These are not well understood.

Recently, Riddle and et al. made an exciting approach to the 
problems mentioned above experimentally \cite{Baker}. By using 
combinatorial chemistry along with a screening strategy, they 
searched and found out a subset of the natural amino acids that 
can be used to construct a well-ordered protein-like molecule 
consisting of $\beta$-sheets. This subset contains five amino 
acids \cite{Baker}, namely, isoleucine, alanine, glycine, glutamic 
acid and lysine, which are simply represented as I, A, G, E and 
K\cite{bch_book}. Although about $30\%$ of the residue sites 
in some of their sequences are not encoded by the five-letter palette, 
rather than by nine other types of residues, the sequence complexity of 
the protein is largely simplified. These uncoded sites are due to 
their direct involvement in binding a proline-rich signal peptide. 
[if the binding was compromised, the protein would not show up in 
the screening array \cite{Wolynes1,Baker}.] They also found that 
subtractions of the five-letter code to a three-letter one may destroy 
the structural rebuilding. As argued by Wolynes\cite{Wolynes1}, 
this experiment shows the possibility of simplification for the 
protein sequence complexity, and that some complexity with five-letter 
code, but not three-letter code, might be still needed based on the 
landscape ideas. This experiment extends the search for minimal 
solution of the simplified representation of complex protein 
architecture. However, it is not clear whether the suggested 5-letter 
code is valid in general and feasible for elucidating characteristics 
of real proteins with 20 kinds of amino acids. That is, is the 
5-letter code based on a specific protein generally workable for 
other natural proteins?

In this work, we address the questions mentioned above at the 
level of a lattice model with some contact-potential-form 
interactions. Based on the statistical and the kinetic 
characteristics of the folding, and on the thermodynamic 
stability of the ground states of some reduced sequences, we 
study the validity of two reductions, namely one with the 
five-letter palette I, A, G, E and K \cite{Baker} and the other 
composed by two types of amino acids as an example of HP-like 
model. We find that the five-letter reduced sequences display 
similar native structural features and folding kinetic behavior 
to the optimal 20-letter ones. Differently, for the two-letter 
case, our results show its deficiency to act as a good 
reduction for general representation of proteins. The five-letter 
reduction may be a suitable description for simplifying the 
sequences complexity of 20-letter model chains which may 
relate to the natural proteins. A detailed analysis on 
correlation between the sequences with 20-letters and their 
five-letter substitutes shows the source of such validity of 
the five-letter code.

\section{MODELS}
In general, a successful reduction implies that an original 
sequence with 20 types of residues can be represented by a 
new sequence with fewer types. The reduced sequence should 
have the main statistical and structural characteristics, 
as well as the kinetic accessibility, as that of the original 
one, such as the basic residue components of the original 
sequence, the ground state or the native conformation, 
etc\cite{Lihao2,TK1,gap1,v}. That is, after 
the reduction, not only should the statistical 
characteristics of the energy spectrum (the existence of a 
large ground state gap) be maintained, but also the folded structure 
should be kept the same as that of the original one (see Fig.1). Such a 
reduction is regarded as a good one. Nevertheless, if the 
energy gap is diminished, or the ground state of the reduced 
sequence is degenerate (or deviates from that of the original 
one), the thermodynamic stability and the kinetic 
characteristics are altered. Consequently, the reduction is 
believed to be a bad one. Therefore, to explore the validity of 
simplified reduction means to verify it being a good one or not. 

Let us start by considering a model protein which is assumed to 
be a chain with connected residues in a cubic lattice 
(Fig.2a). A self-avoided arrangement of the model chain on the 
lattice is generally noted as a structure or conformation of the 
model protein. The conformation of the model chain is 
characterized by the set $\{ {\bf r_i} \} $, the spatial 
position of the $i$-th residue in the chain, and its sequence 
by $\{s_i\}$, the set of various types of residues assigned 
along the chain. An example of such a chain is shown in 
Fig.2(a), with $s_1=V$, $s_2=P$, and $s_3=V$, and so on, and 
the chain length is $L=27$, i.e., $27$ residues. The energy of 
the chain is determined by its sequence and conformation, 
\begin{equation}
   E(\{s_i\},\{ {\bf r_i}\}) = 
   \sum^{L}_{i>j} B(s_i,s_j) \Delta( {\bf r_i}, {\bf r_j}) \, ,
   \nonumber
\end{equation}
where $B(s_i,s_j)$ is the contact potential between the $i$-th 
residue (with residue type $s_{i}$) and the $j$-th residue 
(with residue type $s_j$), and the function 
$\Delta({\bf r_i},{\bf r_j})=1$ when the $i$-th and the $j$-th 
residues are spatial, not sequential, neighbors and 
$\Delta({\bf r_i},{\bf r_j})=0$ otherwise. To study the 
reduction, a number of such sequences, say $N=100$, made of 
the natural set of amino acids with 20 letters are all well 
optimized to a same native structure [the one shown in 
Fig.2(a)] using the methods proposed by 
Shakhnovich et al\cite{design1,design2}. All these sequences 
are referred as ``original'' ones. These original sequences then 
are reduced according to a pre-selected five-letter reduction 
scheme as follows \cite{WW}: group-I (with residues C, M, F, I, 
L, V, W and Y), group-II (A, T and H), group-III (G and P), 
group-IV (D and E) and group-V (S, N, Q, R and K), with a 
representative residue for each group as I, A, G, E and K, 
respectively. That is, an original sequence is reduced into 
a sequence of five-letters by substituting (or replacing) each 
letter with its representative letter. In such a reduction, 
20 types of residues are divided into five groups according to 
their interaction characteristics (following the Miyazawa and 
Jernigan (MJ) matrix\cite{R18a}). Each group contains some 
residues which interact with others in a similar way \cite{WW}. 
For example, group-I includes all the hydrophobic residues, and 
the other four contain the residues with polar features. 
Obviously, this five-group simplification takes into account of 
more heterogeneities for the protein, and also considers more 
detailed differences between the polar residues than the 
two-letter HP model \cite{Wolynes1,WW}. Then we take Baker's 
5-letter alphabet, I, A, G, E and K, as the best representative 
letters for five groups based on a physical reason, but not an 
arbitrary choice (A more detailed description and discussion see 
Ref.\cite{WW,ch}). 

The procedure to reduce the original sequences to two-letter cases 
is similar as above. Here, we consider group $H$ including residues 
C, M, F, I, L, V, W and Y, and group $P$ including the 
rest\cite{HP,Lihao1}, and we pick the residues I and A as the 
representatives of group H and P, respectively. Some other choices, 
such as I and E, are also checked. It is found that these choices 
give out basically the same results.

\section{RESULTS AND DISCUSSIONS}
Does the reduced sequence with five letters have the same folded 
native structure as that of the original one shown in Fig.2(a)? 
Let us make a statistics on the ratio of successful reduction 
($RSR$), for the reduced sequences. Because of the main attractive 
feature of the MJ interaction matrix\cite{R18a}, the native state 
generally has a structural motif with the form of maximally compact 
structure. Besides, for the 27-mer model chain we studied, the 
compact conformations are generally understood as the 
$3\times 3\times 3$ cube-shape 
structures\cite{design2,SaliN,cub3,move}. Thus, the statistics on 
the RSR can be found out by enumerating all possible compact 
structures in the case of a $3\times 3\times 3$ lattice \cite{enum1}. 
Interestingly, almost $86$ reduced sequences (out of $100$) with 
five-letters are {\it ``foldable''} since they have unique ground 
states. Among the foldable sequences, $74$ keep the same native 
structure of the original sequences, termed as ``folded'', and 
$12$ have different folded states, termed as ``ground states 
deviation''  [see Fig.2(b)]. This means that this five-letter strategy 
basically reproduces the original native structure with $RSR$ of $74\%$. 
In addition, there are $14$ sequences that do not have unique ground 
states, termed ``unfoldable''. 

As a comparison, the $RSR$ is also evaluated for the HP reduction. 
It is found that there are a few foldable sequences, namely, three 
reduced sequences (out of $100$) but with ``ground states 
deviation''. That is, these three sequences fold to unique ground 
states which are different from that of the native state of the 
original sequences, i.e., the $RSR$ is $0$. However, the existence of the 
unique ground states basically coincides with the result in 
Ref.\cite{Lihao2} where there is only 4.57\% of the sequences 
having unique ground states. Physically, it results from the 
homopolymeric degeneracy as argued in Ref.\cite{Wolynes1} and 
found in other lattice simulations\cite{HP_bad}. The biochemical 
experiments also approve this deficiency, since it is found that 
two types of residues are far insufficient for the rebuilding of 
four-helix bundle proteins in protein engineering experiments and 
in de novo designs \cite{exper1,exper2}.

Differently, the five-letter case is quite good in overcoming the 
homopolymeric tendency. The similarity between the five-letter case 
and the 20-letter case with a large value of $RSR$ indicates the 
validity of the five-letter strategy in simplifying the natural proteins. 
It is noted that the selection of the optimal original sequences and 
their native structures is rather arbitrary without any bias on 
interactions and compositions. Several studies on some other sequences 
with different native structures present similar results, which further 
implies that this five-letter reduction scheme is universal for different 
structures. Therefore, the simplified depiction of the Riddle et al.'s 
five-letter substitutes may catch basically the physics of 
natural proteins in their structures.

What kinetic characteristics then do the five-letter sequences 
show? These are further studied by the standard lattice Monte 
Carlo simulations\cite{cub3,move}. The thermodynamic quantities, 
such as heat capacity $C_v$ of the model chains, are calculated 
based on the histogram algorithm over a collection of samples of 
conformations\cite{cub3,his}. In our simulations, the native 
structures found in the above enumerations over various compact 
structures are checked. It is found that the checked structures 
have uniquely minimal energies and frequently appear in the sampling 
for these structures. These ensure the thermodynamic stability and 
kinetic accessibility. It approves their characteristics as native 
structures found in our enumerations. Therefore, the statistics on 
the $RSR$ based on these kinetic Monte Carlo simulations shows 
basically the same value five-letter reduction as that obtained by 
enumerations. Similar results have also been obtained for the HP 
reduction.

Furthermore, we use the foldability 
$\sigma=|T_{\theta}-T_f|/T_{\theta}$ proposed by Klimov and 
Thirumalai\cite{TK1} as an indicator of the kinetic 
characteristic for the system. Here $T_{\theta}$ indicates the 
temperature corresponding to the peak of heat capacity of a model 
chain, and $T_f$ marks the structural transition to the well-ordered 
native structure during the folding processes, which corresponds to 
the maximum of the fluctuation of structural overlap function 
$\chi(T)$\cite{TK1}. A small $\sigma$ is generally related 
to a fast folding with few kinetic traps. On the other hand, a large 
$\sigma$ ($\approx 1$) may result in many competitive local minima, 
and in this case a protein can not fold in a biologically relevant 
time scale. Thus, the factor $\sigma$ is argued as a criterion of the 
foldability\cite{TK1} although there is a controversy on 
its definition\cite{sh1}. The correlation between the 
factor $\sigma$ and folding ability has been illustrated in many cases 
(see Ref.\cite{TK} and references therein). Here, we use the 
factor $\sigma$ to characterize the foldability of our model chains.
Note that we do not monitor the actual folding time of the model 
chains in our simulations, and our following discussions on kinetics 
depend on the presumed relationship between the kinetics and 
thermodynamics, i.e., the foldability $\sigma$ proposed in 
Ref.\cite{TK1}. A relevant discussion on such a relationship is 
recently made by Dinner {\it et al} \cite{DP}.

We then calculate the foldability $\sigma$ for the original sequences 
and their five-letter and two-letter substitutes, respectively (see Fig.3, 
the solid circles). One can see that for the five-letter substitutes the 
foldability is $\sigma =0.46$ which is smaller than $0.6$, a critical 
value of slow folding sequences \cite{TK1}. This suggests that 
the native state is accessible kinetically for the five-letter reduction, 
and shows the similar dynamics of two kinds of sequences, namely, the 
original ones and the reduced ones. Differently, for the two-letter case 
the foldability is $\sigma\simeq 1$ much larger than 0.6, which means 
little kinetic accessibility of its native structure. In addition, 
some independently optimized sequences (i.e., some 20-letter sequences 
optimized with the five-letter or two-letter alphabet, respectively) are 
also studied. The values of $\sigma$ are found to be 0.25 and 0.75 for 
the five-letter and two-letter sequences, respectively, which shows clearly 
the improvement of the kinetic accessibility \cite{TK1} due to 
the optimization for each sequence. Moreover, a similar exponential 
decreasing tendency can be seen (see Fig.3, the open diamonds). As a 
result, both exponential tendencies of decrease clearly imply the 
validity of a five-letter strategy.

It is worth noting that in our simulations, a large number of extended 
conformations are explored, not only the most compact conformations within 
the $3\times 3\times 3$ cubic lattice. These kinetic folding simulations
may be important for the case of longer chains since the non-compact 
conformations have more effects on the folding features of the model 
proteins \cite{sh1}. Fortunately, the competition of non-compact 
structure as a native candidate may not be so serious in our case
(see the remarks at the end of this sections). Certainly, for a more 
realistic and detailed study, we may need to make more extensively kinetic 
folding simulations on longer chains, which deserves further work. 
However, we may believe that the basic physics is the same because many main 
characteristics of the folding have been found from the lattice model of 
proteins previously
\cite{Dillchan,Wolynes0,HP,Wolynes1,Lihao1,Lihao2,enum1,TK1}.

Now, let us discuss the correlation between the original sequences 
with 20-letter code and their substitutions with five-letter code. As we 
all know, there is $m^L$ possible sequences for the proteins 
with $m$ types of residues and length $L$. Can the optimized part 
of $20^L$ sequences be successfully mapped to $5^L$ substitutes as 
mentioned-above? To answer this question, we make a detailed 
analyses on the sequence set 
$S^{0}_{20}=\{ S_{20}^1, S_{20}^2, S_{20}^3, \cdots \} $ in which 
all their substitutes with the five-letter reduction are the same, 
i.e., a single sequence $S^0_5$. We randomly select a number of 
compact structures, say 50, from the five-letter spectrum as targets 
for design, i.e., optimizing the sequences $S_{20}^{0}$ to lower 
energy \cite{design1,design2}. 

In practice, from a sequence $S^0_5$, we produce a number of sequences 
$S_{20}^1$, $S_{20}^2$, $\cdots$, $S_{20}^{30}$, randomly following 
the reduction scheme. That is, each of I,A,G,E, and K is randomly 
substituted by a residue belonging to its corresponding group, e.g., 
A by A, T, or H randomly. Then for a certain target structure, we 
process the design for such a sequence $S_{20}^{i}$ by randomly 
exchanging their positions of two residues which belong to the same 
group (keeping the five-letter reduced sequence $S_{5}^{0}$ unchanged). 
The new sequence is accepted when the energy increment $\Delta E <0$, 
otherwise accepted with a probability $P=exp(-\Delta E /T)$ with $T=0.1$. 
Until $10^{7}$ new sequences are reached, the optimized sequence is 
found out for which its energy is the lowest one. The target is 
designable if its energy is the lowest one, otherwise undesignable. 

We find three cases: 
1) Undesignable compact structures (UDS): Some compact structures
never do become the native states of the set $S^{0}_{20}$ even after 
very long time optimization and very slow annealing (see Fig.4). 
2) Designable compact structures with small energy gaps
for their ground states (DSSG):
some compact structures are designable and the related sequences 
fold exactly into these structures. However, the energy 
gaps above their ground states are quite small, generally less 
than 0.5 (see Fig.4). According to the features of the protein energy 
spectrum, such small energy gaps result in an unstable ground state 
\cite{Wolynes0}, and also many traps in the energy 
landscape \cite{Dillchan,Wolynes0}. 
Thus these proteins may fold rather slow and may be unstable 
in their folded states.
3) Designable compact structure with large energy gaps for their 
ground states (DSLG): there exists a specific compact structure, 
namely, the native structure of the sequence $S_5^0$. 
Taking it as a design target, the designed sequences, from 
all of $S^{0}_{20}$, not only fold to it, but also have large 
energy gaps and show good foldability (see Fig.4).
This implies that a set of optimal sequences $S'_{20}$ that have 
the same five-letter reduced sequence shows stable folding and is 
included in the set $S_{20}^0$, $S_{20}'\in S_{20}^{0}$.
Thus, if two optimized sequences $S_{20}'^{i}$ and $S_{20}'^{j}$ 
have the same 5-letter reduced sequence $S_{5}^{0}$, they will behave 
the same in folding behavior as that of $S_{5}^{0}$.
[If sequences with 20 letters are optimized without the condition 
of keeping $S_{5}^{0}$ unchanged (see the preceding discussions), 
most of the optimized sequences (about 74\%) will have the same 
folding behavior after the substitution (see Fig.2(b))]. In other 
words, the complexity of the sequences $S_{20}^{0}$ can be 
simplified via the reduction since both the $S_5^0$ sequence and 
$S_{20}^{0}$ sequences behave almost the same in the aspects of 
folding kinetics and folded structure. Thus by the reduction 
the sequence space (with 20 letters) can be reduced to many sets, 
and each set has a single five-letter sequence and a favorable native 
structure. In this level, the correlation of multiple-to-one between a 
set of 20-letter sequences and a five-letter sequence is well established 
and ensures the similarity of two sequence spaces. Finally, it is 
noted that other five-letter reductions do not present such 
simplification from the original sequences, and have a small (large) 
value of $RSR$ ($\sigma$) since they are not the best representative 
letters for groups \cite{WW}. 

Finally, let us remark on our modeling study. It is argued that 
the enumerations over only the compact structures are questionable 
for searching the native state of a lattice chain\cite{sh1,HP_bad}. 
Indeed, this is true for the usual HP model (with $E_{HH}=-1$ and 
$E_{HP}=E_{PP}=0$) \cite{HP} and that used by Li, et al. (with 
$E_{HH}=-2.3$, $E_{HP}=-1$, and $E_{PP}=0$) \cite{Lihao2}. In those 
models, there are some binding interactions between residues, such as 
$E_{PP}=0$, that are as weak as that between a residue and solvent molecules.
This may mean that there is little difference in energies between the 
compact structures and the non-compact ones in those models 
\cite{sh1}, which introduces some limits in enumeration analysis. 
However, the interactions for our chains, not only for 20-letter chains, 
but also for the five-letter cases and two-letter cases, are taken from the MJ 
matrix directly, which has an obvious attraction between any pairs of 
residues. Thus, there is a large energy penalty for the extended 
structures with fewer bonds, i.e., an energy bias towards maximally 
compact structure. It improves the stability of the compact structures 
in our analysis. This makes our search for the native structures by 
enumerations over the set of compact structures feasible. In addition, 
the lattice chains with 27 monomers are shorter than those analyzed in 
Ref.\cite{HP_bad}. As a result, we believe that in our cases the 
determination of the native structures based on the enumerations over 
compact structures is appropriate, though there is a mixing between 
the energy spectrum (the energy levels above the native state) of the 
compact structures and those of the non-compact structures, which may 
affect the kinetic properties. In practice, the native structures 
obtained by enumerations in our work are verified by the Monte Carlo 
kinetic simulations. That is, for our cases, especially the five-letter 
chains, we have not found any other states with lower energy than the 
native structures during very-long-time simulations. Besides, using 
enumeration over compact structures is more effective than the kinetic 
simulations, which enables us to make an extensive search over the 
sequence space on the structural rebuilding problem, as interested 
in our work. 

\section{SUMMARY}
In this work, the five-letter code for simplification of proteins
suggested by Riddle et al.\cite{Baker} is found to be valid 
not only keeping the features of the energy spectrum but also 
reproducing the native structure of the sequences with the natural 
set of residues. The kinetic resemblance between the sequences 
composed with 20 letters and five-letters also implies a similarity 
of their funnel characteristic. This illustrates that the lattice 
model of proteins with natural set of residues can be re-constructed 
with a smaller set of residues, maybe five types of residues.
Thus our study suggests that the five-letter code can act as a suitable 
simplification considering more heterogeneity of the model proteins, 
and may encode the main information for model proteins. In addition, 
since that the folding kinetics in real proteins and lattice models 
is believed to be similar in many aspects
\cite{Dillchan,Wolynes0,Lihao2}, our simulations on the lattice chains 
may provide some understanding on the simplification of real proteins. 
As a comparison, the unfavorable reductions with two types of residues 
may indicate that the proteins with a natural set of residues are too 
complex to be simplified by the two-letter representations. That is, the 
minimal solution of protein simplification could not be as simple as 
the HP models, at least in lattice level. This actually coincides with 
the conclusion in several protein design experiments
\cite{exper1,exper2,min}. Finally, we note that an exact and 
detailed grouping of residues for the simplification representation 
of proteins still needs more theoretical and experimental explorations.

\section{ACKNOWLEDGMENTS}*
W.W. acknowledges support from the Outstanding Young Research 
Foundation of the National Natural Science Foundation of China
(Grant No. 19625409). We thank C. Tang for a critical reading 
of the manuscript.
The numerical simulations in this work were done on the SGI Origin
2000 in the National Laboratory of Solid State Microstructure, Nanjing
University.

\begin{figure}
Fig.1. Sketchy view of the reduction. For the optimal sequences, 
$S_{20}^1$, $S_{20}^2$ and so on, with 20 types of residues, 
there exist large ground state gaps for their native structures 
in the energy spectrum. A good reduction should retain similar 
spectrum features and keep the native structures (see Substitute I). 
A bad reduction diminishes the energy gap above ground state, and 
the original native structure is not located on lowest energy level. 
\end{figure}

\begin{figure}
Fig.2. (a) A target of folding, and an original optimized 
sequence of L=27 residues (with 20 types) and its five-letter 
and HP substitutes.
(b) Ratio of successful reduction $RSR$ for 100 original 
optimized sequences and its five-letter and two-letter (HP) 
substitutes. The ``folded'', ``ground state deviation'' and  
``unfoldable'' are defined in the text. The interactions 
between the five letters I, A, G, E and K, and those between 
the two letters I and A (or I and E) follow the MJ matrix. 
The results are the same for those two two-letter cases. 
\end{figure}

\begin{figure}
Fig.3. Foldability $\sigma$ versus type number of the reduction
based on the Monte Carlo simulations. The solid circles indicate 
the case of direct substitutes for 50 original optimized sequences, 
and the open diamonds for 50 random sequences optimized with the 
five-letter or two-letter alphabet. The interactions are described in 
Fig.2.
\end{figure}

\begin{figure}
Fig.4. Statistics of design.
(a) The ratio of successful folding to the target (RSF).
    For each target, RSF is obtained by averaging over 
    30 runnings of optimization;  
(b) $\Delta_1$, ground state gaps averaged over all 30 
    runnings of optimization (including the 
    unique and degeneracy ground states).
(c) $\Delta_{2}$, ground state gaps averaged only over 
    the unique ground states. The interactions are 
    described in Fig.2. The  marked positions indicate 
    designs for different structures(see the text).
\end{figure}

\end{multicols}
\end{document}